\documentclass[aps,pra,twocolumn,amsmath,showpacs]{revtex4}
\usepackage{bm}
\usepackage{graphicx}
\begin{document}
\setlength{\arraycolsep}{2pt}
\title{Demonstrating multipartite entanglement of single-particle $W$ states: linear optical schemes}
\author{Hyunchul Nha$^*$ and Jaewan Kim}
\affiliation{School of Computational Sciences, Korea Institute for Advanced Study, Seoul, Korea} 
\date{\today}
%\maketitle
\begin{abstract}
We present two linear optical schemes using {\it nonideal} photodetectors to demonstrate inseparability 
of $W$-type $N$-partite entangled states containing only a single photon. 
First, we show that the pairwise entanglement of arbitrary two modes chosen from $N$ optical modes 
can be detected using the method proposed by Nha and Kim [Phys. Rev. A {\bf 74}, 012317 (2006)], 
thereby suggesting the full inseparability among $N$ parties. 
In particular, this scheme is found to succeed for any nonzero quantum efficiency of photodetectors. 
Second, we consider a quantum teleportation network using linear optics without auxiliary modes. 
The {\it conditional} teleportation can be optimized by a suitable choice of the transmittance of the beam splitter in the Bell measurement. Specifically, we identify the conditions under which maximum fidelity larger than classical bound 2/3 is achieved only in cooperation with other parties. We also investigate the case of on-off photodetectors that cannot discriminate the number of detected photons.  
\end{abstract}
\maketitle
\email{phylove00@gmail.com}

\narrowtext
\section{Introduction}
%\pagenote{*email:phylove00@gmail.com}
Entanglement plays a key role in quantum information processing and it is crucial to characterize entanglement both in theory and in experiment. 
Earlier works on entanglement have been extended to multipartite systems beyond bipartite ones, but our knowledge is still far from complete in many respects. 
Remarkably, it has been shown that only two inequivalent classes of pure states exist in tripartite systems under stochastic local operations and classical communications (SLOCC), namely Greenberger-Horne-Zeilinger (GHZ)-class and $W$-class\cite{Dur}. Some experiments were recently performed to generate and characterize the multipartite entangled states, e.g. the GHZ states\cite{Bouwmeester,Roos}. There have also been several proposals and experimental reports to generate tripartite $W$-states particulary using polarization entangled photons\cite{Eibl,Mikami}, or a trapped-ion system\cite{Roos}. 

In this paper, we will consider a general $N$-partite $W$-states, particularly in the form of {\it mode}-entanglement containing a single photon\cite{Enk}, in contrast to the {\it particle}-entanglement in Ref.\cite{Roos,Eibl,Mikami}. 
We are here concerned with the demonstration of inseparability for the single-particle entangled states using only linear optics with nonideal photodetectors. 
For this purpose, we recall that $W$-states possess a robust two-party entanglement even after all the other $N$-2 parties are traced out\cite{Dur}. 
We first show that the recent proposal by Nha and Kim enables one to detect the two-party entanglement in a simple linear optical setup measuring only a few observables\cite{nha}. Thus, one does not need to employ the quantum tomographic method  as in Refs.\cite{Roos,Eibl,Mikami}, which may be practically demanding in general. 
Particularly, it is shown that the proposed scheme can successfully detect the pairwise entanglement, {\it regardless} of the photodetector efficiency, of {\it arbitrary} two modes chosen from $N$ modes. This proves the genuine multipartite inseparability of the $W$ state. 

Secondly, we consider a linear-optical quantum teleportation without introducing auxiliary modes in a network setting\cite{Yonezawa}. Specifically, we investigate the conditions under which the teleportation fidelity can exceed the classical limit $\frac{2}{3}$\cite{Popescu}. Linear-optical quantum teleportation of vacuum/single-photon qubit, i.e., of the state $a|1\rangle+b|0\rangle$, was previously proposed\cite{Villas,Lee,Knill} and its experimental demonstration also followed\cite{Lombardi}. It is well known that linear optics alone cannot discriminate all the four Bell states so that only a nondeterministic teleportation may be possible\cite{Vaidman,Lutkenhaus}. 
In this paper, we consider a conditional two-party teleportation where Alice and Bob are assisted by $m$ other parties ($0\le m\le N-2)$; The cooperating parties perform measurement in the computational basis to send 1-bit information, and the teleportation protocol proceeds only when these $m$ optical modes are measured in vacuum states. 
We analytically obtain the optimal transmittance of the beam splitter in the Bell measurement and the requirement of photodetector efficiency to achieve fidelity $>\frac{2}{3}$. 
Particularly, we derive the conditions under which the fidelity $>\frac{2}{3}$ can be achieved {\it only in cooperation with other parties}, thereby indicating multipartite entanglement to some extent. 
Furthermore, we will also investigate the case of on-off photodetectors that cannot discriminate the number of detected photons. 

This paper is organized as follows. In Sec.~II, we first present how to prepare an arbitrary $N$-partite $W$-state with a single photon. In Sec.~III, we briefly introduce a separability condition that can be used to detect non-Gaussian entangled states  . Particularly, this condition is refined to include the effect of nonideal photodetectors and then employed to detect arbitrary two-party entanglement in the single-photon $W$-states. 
In Sec.~IV, we introduce a quantum network teleportation and obtain conditions under which the quantum fidelity can exceed the classical limit. In Sec.~V, the main results of this paper are summarized.
 
\section{Generation of single-photon $W$-states}
We first present an optical scheme to prepare an arbitrary $N$-partite $W$-state with a single photon. 
A general $W$-state of the form 
\begin{eqnarray}
|W\rangle=\alpha_1|1,0,\cdots,0\rangle&+&\alpha_2|0,1,\cdots,0\rangle\nonumber\\+\cdots&+&\alpha_N|0,0,\cdots,1\rangle
\label{eqn:w}
\end{eqnarray} 
can be obtained by injecting a single photon into an array of beam splitters as shown in Fig.~1. 
\begin{figure}
\includegraphics*[width=2.2in,keepaspectratio=true]{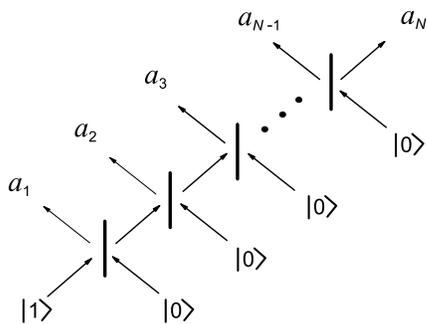}
\caption{Experimental scheme of generating an arbitrary single-particle $W$-state in Eq.~(\ref{eqn:w}). A single-photon at one input is injected into an array of beam splitters with the transmissivity of the $j$-th beam splitter given by $\cos\theta_j$. }
\label{fig:lodi1}
\end{figure}
The $j$-th beam splitter transforms two adjacent modes \{$a_j$, $a_{j+1}$\} into \{$a'_j$, $a'_{j+1}$\} as
\begin{eqnarray}
\begin{pmatrix}&a'_j\\&a'_{j+1}
\end{pmatrix}
=\begin{pmatrix}
&\sin\theta_j&-\cos\theta_j\\&\cos\theta_j&\sin\theta_j
\end{pmatrix}
\begin{pmatrix}&a_j\\&a_{j+1}
\end{pmatrix},
\end{eqnarray}
where $\cos\theta_i$ ($\sin\theta_i$) denote the transmissivity (reflectivity) of the beam splitter\cite{Campos}. 
The coefficients $\alpha_i$ at the output are then given by 
\begin{eqnarray}
\alpha_1&=&\sin\theta_1,\nonumber\\ 
\alpha_2&=&\cos\theta_1\sin\theta_2,\nonumber\\ 
&&\vdots\nonumber\\
\alpha_j&=&\left[\Pi_{i=1}^{j-1}\cos\theta_i\right]\sin\theta_j,\nonumber\\ 
&&\vdots\nonumber\\
\alpha_N&=&\left[\Pi_{i=1}^{N-1}\cos\theta_i\right].
\label{eqn:coeff}
\end{eqnarray}
If a phase shifter is placed in addition at each output with $\phi_j$ the phase shift 
at the $j$-th mode, the coefficients become $\tilde{\alpha}_j=\alpha_je^{-i\phi_j}$ ($i=1,\cdots,N$).

\section{Two-party entanglement}
In this section, we consider two-mode entanglement of the $W$-state in Eq.~(\ref{eqn:w}). 
Two modes $i$ and $j$ arbitrarily chosen are, after the other modes are traced out, reduced to the state   
\begin{eqnarray}
\rho_{ij}=p_{ij}|\Psi_{ij}\rangle\langle\Psi_{ij}|+(1-p_{ij})|00\rangle\langle00|,
\label{eqn:state1}
\end{eqnarray} 
where $|\Psi_{ij}\rangle=\frac{1}{\sqrt{p_{ij}}}\left(a_i|10\rangle+a_j|01\rangle\right)$ and $p_{ij}=|a_i|^2+|a_j|^2$.
The above state is entangled, which can be easily checked by the negativity under partial 
transposition\cite{Peres}, but the issue at hand is how to detect the two-mode entanglement in experiment. 
Note that $\rho_{ij}$ in Eq.~(\ref{eqn:state1}) represents a non-Gaussian state in the viewpoint of continuous variables. 
Recently, several authors obtained separability conditions that can be used to detect entanglement for non-Gaussian states\cite{Hillery,Agarwal,nha}, and it was particularly shown that the inequality derived by Agarwal and Biswas is optimal in a specific class of inequalities based on the su(2) and the su(1,1) algebra involving up to fourth-order moments\cite{nha}. 
The optimal separability condition reads as 
\begin{eqnarray}
\left[1+4(\Delta J_x)^2\right]\left[1+4(\Delta J_y)^2\right]\ge(1+\langle N_+\rangle)^2.
\label{eqn:opt}
\end{eqnarray}
where $J_x=\frac{1}{2}\left(a^\dag b+ab^\dag\right)$, $J_y=\frac{1}{2i}\left(a^\dag b-ab^\dag\right)$, 
and $N_+=a^\dag a+b^\dag b$. 
We will apply this inequality to detect entanglement for $\rho_{ij}$ in Eq.~(\ref{eqn:state1}). 

Particularly, in Ref.\cite{nha}, we proposed a linear optical scheme to test the inequality~(\ref{eqn:opt}) as shown in Fig.~2. 
At the output, one measures the total photon number $N_+=c^\dag c+d^\dag d=a^\dag a+b^\dag b$ and the photon number difference $c^\dag c-d^\dag d$, which becomes $2J_x$ ($2J_y$) for the phase-shift $\phi=0$ ($\phi=\frac{\pi}{2}$). 
We now investigate how such a scheme can yield $\Delta J_x$, $\Delta J_y$ and $N_+$ when the photodetectors have efficiency $\eta\le1$. 
\begin{figure}
\includegraphics*[width=1.8in,keepaspectratio=true]{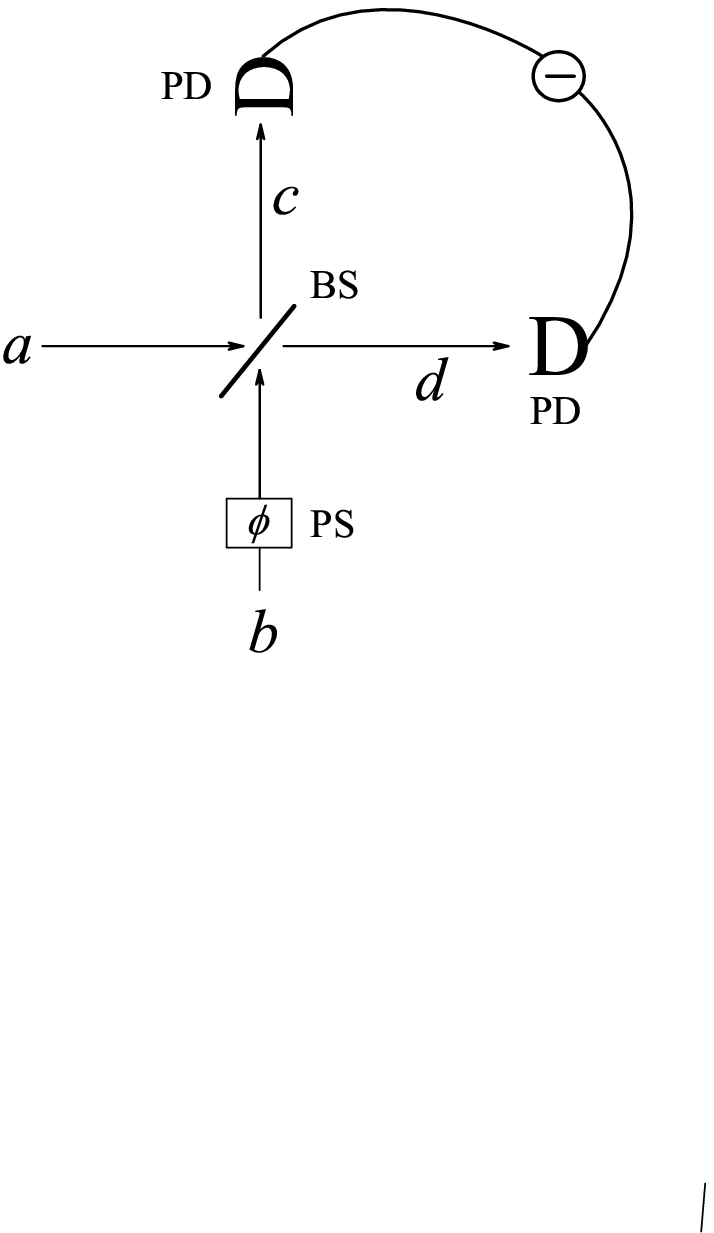}
\vspace{-1.5in}
\caption{Experimental scheme for measuring $J_x=\left(a^\dag b+ab^\dag\right)/2$ and 
$J_y=\left(a^\dag b-ab^\dag\right)/2i$. BS: 50:50 beam-splitter, PS: phase-shifter. $J_x$ ($J_y$) can be detected, 
with the phase shift $\phi=0$ ($\phi=\frac{\pi}{2}$), 
by measuring the photon-number difference $c^\dag c-d^\dag d$ at the output.}
\label{fig:lodi2}
\end{figure}
A nonideal photodetector with quantum efficiency $\eta$ can be modeled as an ideal one but with a beam splitter of transmittance $\eta$ in front of it. 
That is, the actually measured mode $c_{\eta}$ can be expressed as $c_{\eta}=\sqrt{\eta}c+\sqrt{1-\eta}v$, where an auxiliary mode $v$ is in vacuum state. 
The measured variance of number difference is then given by $\Delta^2\left(c^\dag_{\eta}c_{\eta}-d^\dag_{\eta}d_{\eta}\right)=\eta\Delta^2\left(c^\dag c-d^\dag d\right)+\eta(1-\eta)\left(c^\dag c+d^\dag d\right)$. 
In other words, the variances of $J_x$ and $J_y$ are observed in experiment as  
\begin{eqnarray}
4(\Delta J_{i,{\eta}})^2=4\eta(\Delta J_{i,o})^2+\eta(1-\eta)N_{+,o}\hspace{0.5cm}(i=x,y),
\end{eqnarray} 
where the subscript $o$ refers to the original input state. 
The total photon number is of course measured to be $N_{+,{\eta}}=\eta N_{+,o}$. 
Using these relations for the two-mode state $\rho_{ij}$ in Eq.~(\ref{eqn:state1}), we obtain 
\begin{eqnarray}
&&\frac{\left[1+4(\Delta J_{x,{\eta}})^2\right]\left[1+4(\Delta J_{y,{\eta}})^2\right]}{(1+\langle N_{+,{\eta}}\rangle)^2}\nonumber\\
&&=\left(1-\frac{4\eta^2{\rm Re}^2[a_i^*a_j]}{1+\eta p_{ij}}\right)\left(1-\frac{4\eta^2{\rm Im}^2[a_i^*a_j]}{1+\eta p_{ij}}\right).
\label{eqn:fraction}
\end{eqnarray}
The above quantity is always less than unity, which implies the violation of the inequality~(\ref{eqn:opt}), {\it regardless of any parameter values}. 
Therefore, the linear optical scheme in Fig.~2 can detect entanglement of two {\it arbitrary} modes for any nonzero quantum efficiency. 
This means that one can prove the genuine multipartite inseparability of the $W$-state {\it indirectly} in the following fashion. One takes all the different $N(N-1)/2$ pairs of two-modes among $N$ modes and detect the two-party entanglement for each pair [Eq.~(\ref{eqn:fraction})]. 
When one finds the violation of the inequality~(\ref{eqn:opt}) for {\it every} pair, one can conclude that the $W$-state possesses the genuine multipartite inseparability; If, on the other hand, there exists {\it any} separable form possible with respect to different partitions of modes as $\rho=\Sigma_lp_l\rho_{1\cdots i}^{(l)}\otimes\rho_{i+1\cdots j}^{(l)}\otimes\cdots\otimes\rho_{k+1\cdots N}^{(l)}$, there must be at least one pair of modes that are reduced to a separable state, which will contradict our experimental observation.

\section{Quantum network teleportation}
In this section, we discuss a quantum network teleportation via the $W$-state, particularly a symmetric one with $a_j=1/\sqrt{N}$ 
for all $j=1,\cdots,N$ in Eq.~(\ref{eqn:w}), using linear optics {\it without} auxiliary modes. 
The symmetric $W$ state is generated by choosing the reflectivity of the beam splitter as 
\begin{eqnarray}
\sin\theta_j=\frac{1}{\sqrt{N-j+1}} \hspace{0.3cm} (j=1,\cdots,N).
\end{eqnarray} 
In our scheme, two parties, say, Alice and Bob, perform the quantum teleportation of an unknown state $|\phi\rangle=a|1\rangle+b|0\rangle$\cite{Pegg,Pegg1}, and other $m$ parties are in cooperation with them ($0\le m\le N-2)$; $m$ parties carry out measurement in the computational basis $|0\rangle$ and $|1\rangle$ to send one-bit information to Alice and Bob. Only if all the $m$ parties are measured in $|0\rangle$, Alice and Bob agree to do the teleportation. 
Note that if all the other $m=N-2$ parties are measured in vacuum states, Alice and Bob may be in a maximally-entangled Bell state $|\Psi_+\rangle=1/\sqrt{2}\left(|10\rangle+|01\rangle\right)$. With nonideal photodetectors, however, it is likely that no photons are detected even in the presence of a single photon, thereby degrading the entanglement of the remaining two-mode state. 

Detection of $k$-photons at a photodetector with quantum efficiency $\eta$ can be represented by a POVM operator $\Pi_k$ ($\Sigma_k\Pi_k=I$) as
\begin{eqnarray}
\Pi_k=\Sigma_{l\ge k}^\infty\hspace{0.1cm}{}_lC_k\eta^k(1-\eta)^{l-k}|l\rangle\langle l|,
\label{eqn:detector}
\end{eqnarray}
where ${}_lC_k=\frac{l!}{k!(l-k)!}$ counts the number of events where $k$ detected photons are chosen out of $l$ photons\cite{Bartlett}. 
Now the unnormalized two-mode state on the condition that all $m$ parties are measured in $|0\rangle$ is given by 
\begin{eqnarray}
\rho_{AB}^c&=&{\rm tr}_{3\cdots N}\left\{\sqrt{\Pi_0^{\otimes m}}|W\rangle\langle W|\sqrt{\Pi_0^{\otimes m}}\right\}\nonumber\\
&=&\frac{2}{N}|\Psi_+\rangle\langle\Psi_+|+\left(\frac{N-\eta m-2}{N}\right)|00\rangle\langle 00|,
\label{eqn:state_c}
\end{eqnarray}
which is a mixture of the Bell state and the vacuum state. 
We now investigate the quantum teleportation using the conditional state in Eq.~(\ref{eqn:state_c}) and consider first the Bell measurement at Alice's station.

\subsection{Bell measurement}
In the Bell measurement, Alice must distinguish the four Bell states
\begin{eqnarray}
|\Psi_{\pm}\rangle&=&\frac{1}{\sqrt{2}}\left(|10\rangle\pm|01\rangle\right)=\frac{1}{\sqrt{2}}(u^\dag\pm a^\dag)|00\rangle\nonumber\\
|\Phi_{\pm}\rangle&=&\frac{1}{\sqrt{2}}\left(|11\rangle\pm|00\rangle\right)=\frac{1}{\sqrt{2}}(u^\dag a^\dag\pm \hat{I})|00\rangle,
\end{eqnarray}
where $u$ ($a$) are the annihilation operator for the unknown-state mode (Alice's mode). 
In our scheme, a beam splitter that transforms the two modes $u$ and $a$ into $c$ and $d$ as 
\begin{eqnarray}
\begin{pmatrix}&c\\&d
\end{pmatrix}
=\begin{pmatrix}
&\cos\theta&\sin\theta\\&-\sin\theta&\cos\theta
\end{pmatrix}
\begin{pmatrix}&u\\&a
\end{pmatrix},
\label{eqn:trans}
\end{eqnarray}
is used in the Bell measurement and Alice detects photons at the output modes $c$ and $d$. 
Let us denote by $D_{mn}$ the detection event in which $m$($n$) photons are detected at the mode $c$($d$). 
Alice has only six different detection events, i.e., $D_{00},D_{10},D_{01},D_{20},D_{11}$, and $D_{02}$. Among them, only the two events $D_{10}$ and $D_{01}$ are advantageous. For instance, with the choice of $\theta=\pi/4$ in Eq.~(\ref{eqn:trans}), the event $D_{10}$ ($D_{01 }$) corresponds to the Bell state $|\Psi_{+}\rangle$ ($|\Psi_{-}\rangle$) {\it unambiguously}. 
On the other hand, the other events cannot discriminate between $|\Phi_{+}\rangle$ and $|\Phi_{-}\rangle$ for any values of $\theta$. In fact, we can show by an analysis similar to the ones in the next subsections that the teleportation fidelity cannot ever exceed the classical limit 2/3 via the four detection events $D_{00},D_{20},D_{11}$, and $D_{02}$. Therefore, a {\it conditional} scheme to single out only $D_{10}$ and/or $D_{01}$ is favorable to achieve the highest possible fidelity\cite{nha1}. 
We will thus consider only those two events in the following. 

\subsection{case of $D_{10}$ event}
If the detector at mode $c$ ($d$) measures one (zero) photon, the unnormalized Bob's state is given by 
\begin{eqnarray}
\rho_B^{\{10\}}={\rm tr}_{cd}\left\{\sqrt{\Pi_1\otimes\Pi_0}|\phi\rangle\langle\phi|\otimes\rho_{AB}^c\sqrt{\Pi_1\otimes\Pi_0}\right\},
\label{eqn:bob}
\end{eqnarray}
where $|\phi\rangle=a|1\rangle+b|0\rangle$ is an unknown state to teleport 
and $\rho_{AB}^c$ is the conditional state in Eq.~(\ref{eqn:state_c}). 
Note that the POVM operator $\Pi_1\otimes\Pi_0$ in Eq.~(\ref{eqn:bob}) refers to the output modes $c$ and $d$ in Eq.~(\ref{eqn:trans}) after the beamsplitter action. 
We obtain $\rho_B^{\{10\}}$ as 
\begin{eqnarray}
\rho_B^{\{10\}}=\frac{\eta}{N}\left(|\phi'\rangle\langle\phi'|+|a|^2R(\theta)|0\rangle\langle0|\right),
\label{eqn:state_t}
\end{eqnarray}
where 
\begin{eqnarray}
|\phi'\rangle&=&a\cos\theta|1\rangle+b\sin\theta|0\rangle,\nonumber\\
R(\theta)&=&(N-\eta m-2)\cos^2\theta+1-\eta.
\end{eqnarray} 
The probability of $D_{10}$ event is calculated as 
\begin{eqnarray}
P^{\{10\}}&=&{\rm tr}\{\rho_B^{\{10\}}\}\nonumber\\
&=&\frac{\eta}{N}\left(|a|^2\cos^2\theta+|b|^2\sin^2\theta+|a|^2R(\theta)\right).
\label{eqn:prob}
\end{eqnarray}
Note that the state $|\phi'\rangle$ in Eq.~(\ref{eqn:state_t}) is the one left to Bob after the Bell measurement when Alice and Bob initially possess the Bell state $|\Psi_+\rangle$ in Eq.~(\ref{eqn:state_c}). 
We now see that Bob is left with the state $|\phi'\rangle$ or $|0\rangle$ after $D_{10}$ event. 
Bob can only perform a phase-shift in the linear optical scheme and the phase shift does not change the $|0\rangle$ state at all. 
Therefore, to maximize the fidelity, especially $|\langle\phi|\phi'\rangle|^2$, 
Bob performs zero ($\pi$) phase-shift if the transmissivity of Alice's beam splitter is in the range of $\theta\in[0,\frac{\pi}{2}]$ ($[\frac{\pi}{2},\pi]$). 
From now on, it suffices only to consider the case $\theta\in[0,\frac{\pi}{2}]$. 
We now obtain the unnormalized fidelity $P^{\{10\}}\cdot F^{\{10\}}$ as
\begin{eqnarray}
&&P^{\{10\}}\cdot F^{\{10\}}=\langle\phi|\rho_B^{\{10\}}|\phi\rangle\nonumber\\
&=&\frac{\eta}{N}\left[\left(|a|^2\cos\theta+|b|^2\sin\theta\right)^2+|a|^2|b|^2R(\theta)\right].
\label{eqn:fidelity}
\end{eqnarray} 
We will now suppress the use of the superscript $\{10\}$ in this subsection. 
Finally, we consider as a figure of merit the fidelity averaged over all possible input states, that is, 
$\bar{F}=\frac{\int d\mu PF}{\int d\mu P}$, where $\int d\mu=\frac{1}{4\pi}\int \sin\theta_{\rm i}d\theta_{\rm i}d\phi_{\rm i}$ denotes the average over the input states in the Bloch sphere 
with $a=\cos\frac{\theta_{\rm i}}{2}e^{-i\phi_{\rm i}}$ and $b=\sin\frac{\theta_{\rm i}}{2}$. 
%with $|a|^2=\mu$ 
%and $|b|^2=1-\mu$ uniformly distributed in [0,1]. (Note that the average over the relative phase of $|1\rangle$ and 
%$|0\rangle$ states are irrelevant because $P$ and $PF$ in Eqs.~(\ref{eqn:prob}) and~(\ref{eqn:fidelity}) depend only on the moduli of the complex coefficients $a$ and $b$.)
The averaged fidelity $\bar{F}$ is then given by 
\begin{eqnarray}
\bar{F}_{N,m}(\theta)=\frac{1}{3}\left[1+\frac{1+\sin2\theta}{1+R(\theta)}\right],
\label{eqn:fidelity_a}
\end{eqnarray}
and the averaged success probability $\bar{P}=\int d\mu P$  by
\begin{eqnarray}
\bar{P}_{N,m}(\theta)=\frac{\eta}{2N}\left[1+R(\theta)\right].
\label{eqn:prob_a}
\end{eqnarray}
The fidelity in Eq.~(\ref{eqn:fidelity_a}) reaches a maximum value 
\begin{eqnarray}
\bar{F}_{N,m}^{\rm max}=\frac{1}{3}\left[1+\frac{N-\eta (m+2)+2}{(2-\eta)(N-\eta m-\eta)}\right].
\label{eqn:fidelity_m}
\end{eqnarray}
for the choice of the beam splitter transmissivity at Alice's station as 
\begin{eqnarray}
\cos\theta=\frac{(2-\eta)}{\sqrt{(2-\eta)^2+(N-\eta m-\eta)^2}}.
\label{eqn:choice}
\end{eqnarray}
For such a choice, the averaged probability is given by 
\begin{eqnarray}
\bar{P}_{N,m}=\frac{\eta(2-\eta)}{2N}\left[1+\frac{(2-\eta)(N-\eta m-2)}{(2-\eta)^2+(N-\eta m-\eta)^2}\right].
\label{eqn:prob_m}
\end{eqnarray}
Eqs.~(\ref{eqn:fidelity_m}),~(\ref{eqn:choice}), and~(\ref{eqn:prob_m}) are the main results of Section IV, and we will discuss them in Sec.~IV~E.

\subsection{case of $D_{01}$ event}
In the case of $D_{01}$ event, all the previous results in Sec.~IV~B remain the same, but only with the replacement of $\theta\rightarrow\theta+\pi/2$. 
In addition, Bob must now perform $\pi$ phase-shift when Alice's beam splitter is in the range of $\theta\in[0,\frac{\pi}{2}]$.

\subsection{Combination of $D_{10}$ and $D_{01}$ events}
One can also consider the conditional teleportation scheme to include both the events $D_{10}$ and $D_{01}$. We then obtain the maximum of the averaged fidelity $\bar{F}=\frac{\int d\mu \left(P^{(10)}F^{(10)}+P^{(01)}F^{(01)}\right)}{\int d\mu \left(P^{(10)}+P^{(01)}\right)}$ as 
\begin{eqnarray}
\bar{F}_{N,m}^{\rm max}=\frac{1}{3}\left[1+\frac{4}{N-\eta (m+2)+2}\right],
\label{eqn:fidelity_mc}
\end{eqnarray}
for the choice of $\theta=\frac{\pi}{4}$, and the averaged success probability $\bar{P}=\int d\mu \left(P^{(10)}+P^{(01)}\right)$ as
\begin{eqnarray}
\bar{P}_{N,m}=\frac{\eta}{2N}\left[N-\eta (m+2)+2\right].
\label{eqn:prob_mc}
\end{eqnarray}
It turns out that the fidelity in Eq.~(\ref{eqn:fidelity_mc}) is always smaller than or equal to that in Eq.~(\ref{eqn:fidelity_m}). 
Therefore, so long as the goal is to attain the maximum possible fidelity even at the expense of success probability, 
the conditional scheme to single out only one event $D_{10}$ or $D_{01}$ becomes the best one\cite{nha1}. 
Note that for $N=2$ ($m=0$) with ideal photodetector $\eta=1$, the fidelity $\bar{F}_{2,0}$ becomes unity 
with the success probability $\bar{P}=1/2$ from Eqs.~(\ref{eqn:fidelity_mc}) and~(\ref{eqn:prob_mc}), which recovers the case previously considered in Refs.\cite{Lee,Knill}. 

\subsection{Fidelity $>\frac{2}{3}$}
Let us investigate under what conditions the maximum fidelity in Eq.~(\ref{eqn:fidelity_m}) can be larger than the classical limit $\frac{2}{3}$. 
For a fixed $N$, the fidelity monotonously increases as the cooperation number $m$ increases, i.e., it is always advantageous to have more parties in cooperation with Alice and Bob for any values of $\eta$. 
If the photodetectors are all ideal, $\eta=1$, the quantum fidelity in Eq.~(\ref{eqn:fidelity_m}) becomes $\bar{F}_{N,m}^{\rm max}=\frac{1}{3}\left[1+\frac{N-m}{N-m-1}\right]>\frac{2}{3}$ for any $m$, that is, no cooperation is needed to overcome the classical bound. 

In general, the quantum efficiency $\eta$ must be large enough, $\eta>\eta^c$, to have fidelity $>\frac{2}{3}$, and we obtain the critical efficiency $\eta^c$ as 
\begin{eqnarray}
\eta_{N,m}^c=\frac{N+m-\sqrt{(N-m-2)^2+4(m+1)}}{2(m+1)}.
\label{eqn:eta_c}
\end{eqnarray}  
Let us now consider two cases of small number $N$.\\
(i) case of $N=2$ (maximally entangled EPR pair). From Eqs.~(\ref{eqn:fidelity_m}) and~(\ref{eqn:choice}), we have $\bar{F}_{2,0}^{\rm max}=\frac{1}{3}(1+\frac{2}{2-\eta})>\frac{2}{3}$ at $\theta=\frac{\pi}{4}$ for {\it any} $\eta>0$, therefore the manifestation of quantum entanglement is always successful.\\
(ii) case of $N=3$. In this case, from Eq.~(\ref{eqn:eta_c}), we obtain $\eta_{3,0}^c=\frac{3-\sqrt{5}}{2}\approx0.382$, and 
$\eta_{3,1}^c=\frac{2-\sqrt{2}}{2}\approx0.293$. Thus, the quantum efficiency $\eta$ of the photodetectors must be at least larger than $\eta_{3,1}^c\approx0.293$ to achieve $\bar{F}^{\rm max}>\frac{2}{3}$. In particular, for $\eta\in(\eta_{3,1}^c,\eta_{3,0}^c)$, 
the quantum fidelity can exceed $\frac{2}{3}$ only when the third party is in cooperation with Alice and Bob, thereby manifesting tripartite entanglement.   

In general, the critical value $\eta_{N,m}^c$ increases as a function of $N$, and for a fixed $N$, it decreases with increasing $m$. 
That is, the requirement of quantum efficiency becomes less demanding as more parties are in cooperation with Alice and Bob.
In particular, we have $\eta_{N,N-2}^c=1-\frac{1}{\sqrt{N-1}}$ and $\eta_{N,N-3}^c=\frac{2N-3-\sqrt{4N-7}}{2N-4}$. 
When $\eta$ lies in the range of $\eta\in\left(\eta_{N,N-2}^c,\eta_{N,N-3}^c\right)$, the teleportation fidelity can exceed the classical limit only when all the other $N-2$ parties are in cooperation, which thus manifests $N$-partite entanglement. Of course, this interpretation may hold good in somewhat limited context: 
Instead of seeking other parties' cooperation, Alice and Bob may choose to purify their mixed state to a maximally entangled one\cite{Bennett}, or to adopt an error-correcting teleportation scheme\cite{Knill}. 
Such schemes, however, usually require more resources, and we have restricted our attention in this paper to linear optics without auxiliary modes.

\subsection{On-off photodetectors}
In the previous sections, we assumed that the photodetectors might be able to count the number of photons despite the nonideal efficiency $\eta<1$. 
If, on the other hand, the detectors can only discriminate two events, namely, no-photon or some photons, the POVM operators corresponding to these events are given by $\Pi_0$ and $\Pi_{\rm s}\equiv I-\Pi_0$, where $\Pi_0$ is given by Eq.~(\ref{eqn:detector}). 
Now, the detection event $D_{{\rm s}0}$, instead of $D_{10}$ previously considered, is represented by the POVM operator $\Pi_{\rm s}\otimes\Pi_0$, and after some calculation, 
we find the averaged fidelity as
\begin{eqnarray}
\bar{F'}_{N,m}(\theta)
=\frac{1}{3}\left[1+\frac{1+\sin2\theta}{1+R(\theta)+R'(\theta)}\right].
\label{eqn:fidelity_aa}
\end{eqnarray}
and the averaged success probability as 
\begin{eqnarray}
\bar{P}_{N,m}(\theta)=\frac{\eta}{2N}\left[1+R(\theta)+R'(\theta)\right].
\label{eqn:prob_aa}
\end{eqnarray}
We see that a new term $R'(\theta)\equiv2\eta\sin^2\theta\cos^2\theta$ appear in the above equations [Cf. Eqs.~(\ref{eqn:fidelity_a}) and~(\ref{eqn:prob_a})]. Therefore, the fidelity is smaller than that in the previous case, $\bar{F'}_{N,m}<\bar{F}_{N,m}$, and the critical efficiency of photodetectors are increased. We can maximize the fidelity in Eq.~(\ref{eqn:fidelity_aa}) numerically by adjusting the parameter $\theta$ and obtain the critical efficiencies, e.g., for $N=3$, as ${\eta'}_{3,1}^c\approx0.435$ and ${\eta'}_{3,0}^c\approx0.583$. [Cf. Sec.~IV E] 

Note, however, that the result of Sec.~III is unchanged even in the case of on-off photodetectors, because the $W$-state in Eq.~(\ref{eqn:w}) has at most one single photon.

\section{Summary} 

In this paper, we presented two linear optical schemes using nonideal photodetectors to detect inseparability of single-particle $W$-state. 
In the first scheme, it was shown that arbitrarily chosen two-party entanglement can be detected, thereby proving multipartite inseparability, {\it regardless of photodetector efficiency}. 
In the second scheme of conditional teleportation based on a network setting, we investigated the experimental conditions under which the teleportation fidelity can be made larger than the classical limit 2/3 assisted by other parties, which manifests multipartite entanglement to some extent. 
We also discussed the effect of using on-off photodetectors in both of the schemes. 
These schemes seem to be experimentally realizable within the current technology, 
considering particularly the experimental achievement in Refs.\cite{Lombardi} and \cite{Pegg1}.

We acknowledge the financial support from the Korea Ministry of Information and Communication.\\

*email:phylove00@gmail.com


\begin{references}
\bibitem{nha1} This conditional scheme, of course, may not be very useful as a teleportation tool, 
but note that our aim in this paper is to demonstrate quantum entanglement of the $W$-state. 
The quantum fidelity over 2/3 is still a manifestation of quantum entanglement because it cannot be achieved using only classical resources even in conditional schemes.
\end{references}
\end{document}